# The Shortcomings of Video Conferencing Technology, Methods for Revealing Them, and Emerging XR Solutions

**Dani Paul Hove** [1] **and Benjamin Watson** [1]

[1]*Visual Experience Lab, North Carolina State University, Department of Computer Science, School of Engineering, Raleigh, North Carolina, USA*

Correspondence*:
Dani Paul Hove
dphove@ncsu.edu

**ABSTRACT**

Video conferencing has become a central part of our daily lives, thanks to the COVID-19 pandemic. Unfortunately, so have its many limitations, resulting in poor support for communicative and social behavior and ultimately, "zoom fatigue." New technologies will be required to address these limitations, including many drawn from mixed reality (XR). In this paper, our goals are to equip and encourage future researchers to develop and test such technologies. Toward this end, we first survey research on the shortcomings of current video conferencing systems. We then consider the methods that research uses to evaluate support for communicative behavior, and argue that those same methods should be employed in identifying, improving and validating promising video conferencing technologies. Next we survey emerging XR solutions to video conferencing's limitations, most of which do not employ head-mounted displays. We conclude by identifying several opportunities in video conferencing research.

## 1 INTRODUCTION

Over the past two years of the COVID-19 pandemic, a mass move toward remote work and communication has forced many to reckon with the long term effects of video conferencing as a primary communication method. In particular, Zoom fatigue, or video conferencing fatigue, has become particularly prominent (Fauville et al., 2021a). Zoom fatigue is defined as physical and cognitive exhaustion resulting from intensive use of video conferencing tools (Riedl, 2021). Post-pandemic, most expect remote video conferencing to remain much more widely used than it was before COVID (Remmel, 2021), serving as both a safety precaution and a crucial enabler of a burgeoning hybrid work environment. Given this, understanding the challenges and opportunities of video conferencing is particularly important, both to prevent negative consequences, and to realize benefits in the long term. This is especially pertinent as use of previously unconventional meeting environments, such as virtual reality, grows.

Toward this end, we survey research and opinion on video conferencing across several disciplines — both technical and social — and across several technologies, including traditional interfaces as well as mixed and virtual reality. We focus in particular on video conferencing's shortcomings and evaluative methods for finding them. Finally, we survey the limited number of emerging solutions that address known shortcomings, including several in VR and XR.

## 2 SURVEY METHODS

Our survey fits within the narrative review framework, defined as a qualitative method seeking to describe current literature, without quantitative synthesis (Shadish et al., 2001). Given the limited literature on video conferencing over the past decades and the recent drastic increase in its use, video conferencing research is still nascent. For this reason, we turned to this narrative review method, which "focuses on formulating general relations among a number of variables of interest". Two examples of this sort of survey include Lam et al. (2011) and Perer and Schneiderman (2009).

We sought to survey references describing video conferencing systems and challenges, as well as the human behavior they sought to support and methods for evaluating that support. To collect references, we therefore first searched Google Scholar for material using the keywords "zoom fatigue", "fatigue", "video conferencing" or "gaze awareness". For references older than 2015, we set a threshold of 10 citations for acceptability, to restrict our survey to work that had had at least minimal impact, when there had been time for the research community to respond. The initial search returned approximately 180 papers. Both authors then examined each paper, discarding any that they agreed did not discuss video conferencing, communicative behavior, or methods for measuring and evaluating that behavior. We then studied the bibliographies of each remaining paper, examining any cited paper with a title that contained our search keywords, and discarding any that we agreed did not meet our inclusion criteria. We were left with 65 papers.

We next categorized these references using an iterative open coding (Creswell, 2014) methodology that began with three labels reflecting our concerns as we began our survey work: *Measures*, or methods for evaluating video conferencing success; *Shortcomings*, or weaknesses of video conferencing solutions; and *Fatigue,*, or the general feeling of exhaustion that many report feeling after video conferencing. Within the *Measures* label, our sub-labels were *technical*, objective methods of measurement; *behavioral* observational and experimental methods capturing human use; and *subjective* methods that asked users to offer their judgements of video conferencing systems. Within the *Shortcomings* label, our sub-labels were *delay*, the lag users encounter between the moment they act and the moment that act is displayed to other conference participants; and *gaze*, the degree to which a system communicates where users are looking. Finally with the *Fatigue* label, we used a *zoom* sub-label to mark discussion of fatigue in the context of video conferencing, and a *general* sub-label to indicate discussion of fatigue more generally.

As we iterated through the papers we found, we proposed additional labels and sub-labels, and adopted them if both authors approved. To the *Shortcomings* label, we added an *objects of discussion* sub-label referring to video conferencing support for referencing items of participant interest, and a *non-verbal cues* sub-label marking support for non-verbal communicative signals beside gaze and discussed objects. We also added a *Focus* label, referencing the type of knowledge a paper contains, with the sub-labels *solution* for an engineering improvement to video conferencing systems, *new measures* for methods of measuring communicative behavior; *review* for a survey of video conferencing research or communicative behavior, and *study* for an experiment examining communication in conferencing systems.

Table 1 shows how we categorized the papers in this survey with our labels. We use these categories to structure the following review.

**Table 1.** *A table categorizing the research reviewed in this paper, by type of evaluation measure, video conferencing shortcoming, type of fatigue, and whether the paper describes new solutions, new evaluative measures, or is a survey of other work.*

|  | Measures | | | Shortcomings of Video Conferencing | | | |
|---|---|---|---|---|---|---|---|
| Authors | Technical | Behavioral | Subjective | Delay | Gaze | Objects of Discussion | Non-Verbal Cues |
| Aaronson et. al. | | | | | | | |
| Angelopoulos et. al. | ■ | | | | ■ | | |
| Bailenson (2020) | | | | | | | |
| Bailenson (2021) | | | | | | | |
| Becher et. al. | ■ | ■ | ■ | ■ | | | ■ |
| Bennett et. al. | | ■ | ■ | | | | ■ |
| Berndtsson et. al. | | | ■ | | | | ■ |
| Bohannon et. al. | | | | | ■ | | |
| Bos et. al. | | ■ | ■ | | ■ | | ■ |
| Boyaci et. al. | ■ | | | ■ | | | |
| D'Angelo and Begel | | ■ | | | ■ | ■ | |
| D'Angelo and Gergle | | ■ | ■ | | ■ | ■ | |
| deHahn | | | | | | | |
| Driskell and Radtke | | ■ | | | | | ■ |
| Driskell et. al. | | ■ | | | | | ■ |
| Edelmann et. al. | ■ | | | | | ■ | ■ |
| Fauville et. al. (2021a) | | | ■ | | | | |
| Fauville et. al. (2021b) | | | ■ | | | | ■ |
| Feick et. al. | | ■ | ■ | | | ■ | |
| Friston & Steed | ■ | | | ■ | | | |
| Gan et. al. | | ■ | | | ■ | | ■ |
| Gasteratos et. al. | | | ■ | | | | ■ |
| Gergle et. al. | | ■ | | | | ■ | |
| Grayson and Monk | | ■ | | | ■ | | |
| Gunkel et. al. | ■ | | ■ | ■ | | | |
| He et. al. (2021) | | | ■ | ■ | ■ | | |
| He et. al.(2020) | | ■ | ■ | | | ■ | ■ |
| Homaeian et. al. | ■ | | | ■ | | | ■ |
| Hopkins and Benford | | | | | | ■ | |
| Ishibashi et. al. | ■ | | ■ | ■ | | | |
| ITU (2000) | ■ | | ■ | | | | |
| ITU (2007) | ■ | | | | | | ■ |
| Jansen & Bulterman | ■ | | | ■ | | | |
| Jansen et. al. | ■ | ■ | ■ | | | | ■ |
| Kobayashi et. al. | | ■ | | | ■ | | |
| Kraut et. al. | ■ | | | ■ | | ■ | |
| Kuster et. al. | ■ | | | | | | |
| Lawrence et. al. | | ■ | ■ | | ■ | | ■ |

|  | Measures | | | Shortcomings of Video Conferencing | | | |
|---|---|---|---|---|---|---|---|
| **Authors** | **Technical** | **Behavioral** | **Subjective** | **Delay** | **Gaze** | **Objects of Discussion** | **Non-Verbal Cues** |
| Li. et. al. | ✓ |  |  |  | ✓ |  |  |
| Lombard et. al. |  |  | ✓ |  |  |  | ✓ |
| Monk and Gale |  | ✓ | ✓ |  | ✓ |  |  |
| Mota and Pimenta |  |  | ✓ |  |  |  |  |
| Nardi and Whittaker |  |  | ✓ |  |  |  | ✓ |
| Nesher Shoshan and Wehrt |  | ✓ | ✓ |  |  |  | ✓ |
| Nguyen & Canny |  |  |  |  | ✓ | ✓ |  |
| Nguyen et. al. |  |  | ✓ |  |  |  | ✓ |
| O'Malley et. al. | ✓ |  |  | ✓ |  |  | ✓ |
| Oducado et. al. |  |  | ✓ |  |  |  | ✓ |
| Olbertz-Siitonen |  | ✓ |  | ✓ |  |  |  |
| Orlosky et. al. |  |  |  |  |  | ✓ | ✓ |
| Pachnowski |  |  |  |  |  | ✓ |  |
| Ratan et. al. |  |  | ✓ |  |  |  |  |
| Riedl |  |  |  |  |  |  | ✓ |
| Roberts et. al. | ✓ |  | ✓ | ✓ |  | ✓ |  |
| Schmitt et. al. | ✓ |  |  | ✓ |  |  | ✓ |
| Schoenenberg |  | ✓ | ✓ | ✓ |  |  | ✓ |
| Schoenenberg et. al. |  |  | ✓ | ✓ |  |  | ✓ |
| Stotts et. al. | ✓ |  | ✓ |  | ✓ | ✓ |  |
| Su et. al. | ✓ |  |  | ✓ | ✓ | ✓ | ✓ |
| Tam et. al. |  | ✓ | ✓ |  |  |  | ✓ |
| Vertegaal et. al. | ✓ | ✓ |  |  | ✓ |  |  |
| Whittaker et. al. |  | ✓ |  |  | ✓ |  | ✓ |
| Wu et. al. |  |  |  |  | ✓ | ✓ | ✓ |
| Yan et. al. |  |  | ✓ |  |  |  | ✓ |
| Yoshimura and Borst |  |  | ✓ |  |  | ✓ | ✓ |
| Yu et. al. |  | ✓ | ✓ |  |  | ✓ |  |

|  | Fatigue | | Focus | | | |
|---|---|---|---|---|---|---|
| **Authors** | **Zoom Fatigue** | **General Fatigue** | **Solution** | **New Measures** | **Review** | **Study** |
| Aaronson et. al. |  | ✓ |  |  | ✓ |  |
| Angelopoulos et. al. |  |  | ✓ |  |  |  |
| Bailenson (2020) | ✓ |  |  |  |  |  |
| Bailenson (2021) | ✓ |  |  |  |  |  |
| Becher et. al. |  |  |  |  |  | ✓ |

|  | Fatigue | | Focus | | | |
| --- | --- | --- | --- | --- | --- | --- |
| Authors | Zoom Fatigue | General Fatigue | Solution | New Measures | Review | Study |
| Bennett et. al. | ● |  |  |  |  | ● |
| Berndtsson et. al. |  |  |  | ● |  | ● |
| Bohannon et. al. |  |  |  |  | ● |  |
| Bos et. al. |  |  |  |  | ● |  |
| Boyaci et. al. |  |  |  | ● |  |  |
| D'Angelo and Begel |  |  | ● |  |  |  |
| D'Angelo and Gergle |  |  | ● |  |  |  |
| deHahn | ● |  |  |  |  |  |
| Driskell and Radtke |  |  |  |  |  | ● |
| Driskell et. al. |  |  |  |  | ● |  |
| Edelmann et. al. |  |  | ● |  |  |  |
| Fauville et. al. (2021a) | ● |  |  | ● |  |  |
| Fauville et. al. (2021b) | ● |  |  |  |  | ● |
| Feick et. al. |  |  | ● |  |  |  |
| Friston & Steed |  |  |  | ● |  |  |
| Gan et. al. |  |  |  |  |  | ● |
| Gasteratos et. al. | ● |  |  |  |  | ● |
| Gergle et. al. |  |  |  |  |  | ● |
| Grayson and Monk |  |  |  |  |  | ● |
| Gunkel et. al. |  |  |  |  |  | ● |
| He et. al. (2021) |  |  | ● |  |  |  |
| He et. al.(2020) |  |  | ● |  |  |  |
| Homaeian et. al. |  |  |  | ● |  |  |
| Hopkins and Benford |  |  | ● |  |  |  |
| Ishibashi et. al. |  |  |  |  |  | ● |
| ITU (2000) |  |  |  | ● |  |  |
| ITU (2007) |  |  |  | ● |  |  |
| Jansen & Bulterman |  |  |  | ● |  |  |
| Jansen et. al. |  |  | ● |  |  |  |
| Kobayashi et. al. |  |  | ● |  |  |  |
| Kraut et. al. |  |  |  |  |  | ● |
| Kuster et. al. |  |  | ● |  |  |  |
| Lawrence et. al. |  |  | ● |  |  |  |
| Li. et. al. |  |  | ● |  |  |  |
| Lombard et. al. |  |  |  | ● |  |  |
| Monk and Gale |  |  | ● |  |  |  |
| Mota and Pimenta |  | ● |  | ● |  |  |
| Nardi and Whittaker |  |  |  |  |  | ● |

|  | Fatigue | | Focus | | | |
| --- | --- | --- | --- | --- | --- | --- |
| Authors | Zoom Fatigue | General Fatigue | Solution | New Measures | Review | Study |
| Nesher Shoshan and Wehrt | ■ | | | | | ■ |
| Nguyen & Canny | | | ■ | | | |
| Nguyen et. al. | | ■ | | | | ■ |
| O'Malley et. al. | | | | | | |
| Oducado et. al. | ■ | | | | | ■ |
| Olbertz-Siitonen | | | | | | ■ |
| Orlosky et. al. | | | ■ | | ■ | |
| Pachnowski | | | | | | |
| Ratan et. al. | ■ | | | | | ■ |
| Riedl | ■ | | | | ■ | |
| Roberts et. al. | | | | | | ■ |
| Schmitt et. al. | | | | | | ■ |
| Schoenenberg | | | | | | ■ |
| Schoenenberg et. al. | | | | | | ■ |
| Stotts et. al. | | | ■ | | | |
| Su et. al. | | | ■ | | | |
| Tam et. al. | | | | | | ■ |
| Vertegaal et. al. | | | ■ | | | |
| Whittaker et. al. | | | | | ■ | |
| Wu et. al. | | | | | | ■ |
| Yan et. al. | | | | ■ | | |
| Yoshimura and Borst | | | | | | ■ |
| Yu et. al. | | | ■ | | | |

## 3  (ZOOM) FATIGUE

The pandemic has drastically increased use of video conferencing, resulting in the widespread experience of what has come to be called "Zoom fatigue". In this section, we first review recent popular literature addressing Zoom fatigue. Next, we investigate research literature on the general phenomenon of fatigue (Table 1, Fatigue, General Fatigue column), and Zoom fatigue itself (Fatigue, Zoom Fatigue column): specifically, its definition, measurement and analyses of its components.

### 3.1  Recent Expert and Popular Opinion

The massive increase in use of video conferencing during the pandemic created a rush of opinion about video conferencing's shortcomings, many focusing on apparent long term consequences. The phrase "Zoom fatigue" was quickly coined (Sklar, 2020; Fosslien and Duffy, 2020; Degges-White, 2020; deHahn, 2020; Rosenberg, 2020; Robert, 2020), referring to a lasting fatigue born from the unique stresses of remote work using video conferencing (this phrase has become a standard, despite the existence of many video conferencing alternatives). A number of possible causes have been suggested. These include *physical issues*, related to the bad ergonomics of turning a freely moving, immersive meeting into a constrained event passing through a small rectangle (Degges-White, 2020); and *emotional issues*, like the turmoil and isolation of a pandemic (Degges-White, 2020). But most thinking has dwelled on *non-verbal and social cues*, including poor eye contact and seeming inattentiveness (Degges-White, 2020), "constant gaze" with a gallery of other faces creating the perception of being watched (Rosenberg, 2020; Jiang, 2020) (Bailenson, 2020), deficiency of gesture that requires a draining hyper-focus to pick up the little body language that remains (Hickman, 2020; deHahn, 2020), "big face," in which faces appear larger than they would at a comfortable interpersonal distance (Bailenson, 2021), and poor backchanneling, which makes it harder to recover from misunderstandings and attentional lapses (Fosslien and Duffy, 2020; Sklar, 2020) through asides with other participants. Appropriately, much of this popular conjecture has been investigated through formal research.

### 3.2 Defining and Measuring Zoom Fatigue

Fatigue is "an unpleasant physical, cognitive, and emotional symptom described as a tiredness not relieved by common strategies that restore energy. Fatigue varies in duration and intensity, and it reduces, to different degrees, the ability to perform the usual daily activities" (Aaronson et al., 1999). When measuring subjective fatigue, "(1) There was found to be a high correlation between the frequency of complaints of fatigue and the feeling of fatigue. (2) The amount of feeling of fatigue is different for the type of symptom." (Yoshitake, 1971). These considerations should apply to Zoom fatigue, which is currently measured subjectively (Nesher Shoshan and Wehrt, 2021).

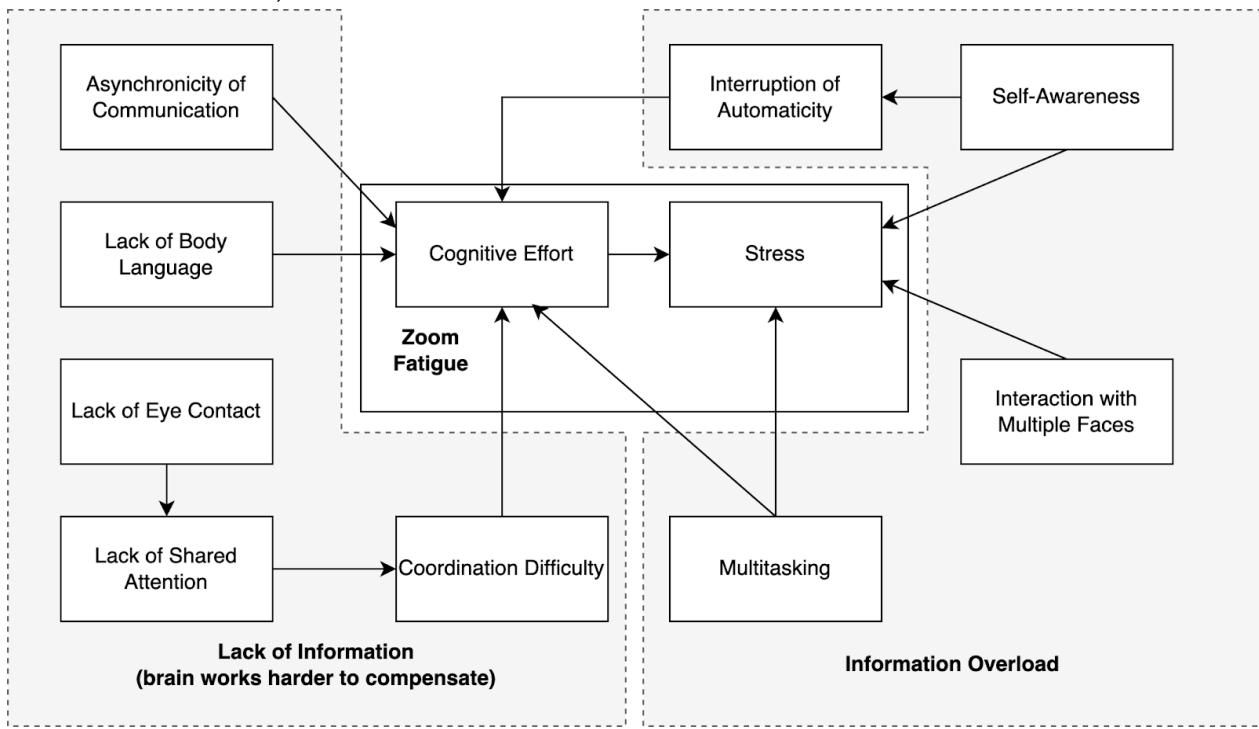

**Figure 1.** *A diagram depicting Riedl's hypothesis for a conceptual model of what factors contribute to Zoom Fatigue, adapted from Riedl's study (2021)*

Riedl et al. (2021) define "Zoom fatigue" as "somatic and cognitive exhaustion that is caused by the intensive and/or inappropriate use of video conferencing tools, frequently accompanied by related symptoms such as tiredness, worry, anxiety, burnout, discomfort, and stress, as well as other bodily symptoms such as headaches." Riedl et al. also posit a conceptual framework for Zoom fatigue (Figure 1), in which both a lack of information (asynchronicity of communication, coordination difficulty, lack of body language, eye contact and shared attention) and an information overload (self awareness or constant gaze, interruption of automaticity, interaction with multiple faces) contribute to cognitive effort and stress. Recently, Fauville et al. (2021a) established a survey measure for Zoom exhaustion, the Zoom Exhaustion and Fatigue (ZEF) Scale, which consists of 15 items across five dimensions of fatigue: general, social, emotional, visual and motivational.

### 3.3 The Components of Zoom Fatigue

Video conferencing delivers several types of information not present in face-to-face meetings, creating a stressful information overload. Much of this information is delivered non-verbally, as indicated by Figure 1. One study suggests that nostalgia for a time before the pandemic may contribute to Zoom fatigue (Nesher Shoshan and Wehrt, 2021), but most thought points toward causes present prior to the COVID-19 crisis. The prominence of self-video in video conferencing has led to a rise in facial dissatisfaction, or mirror anxiety (Fauville et al., 2021b), with some cases extending into "Zoom dysmorphia," driving an increase in plastic surgery (Gasteratos et al., 2021) — particularly among women (Ratan et al., 2021). Hyper-gaze from a grid of staring faces is yet another informational challenge (Fauville et al., 2021b). On the other hand, much of the information normally present in-person is missing in video conferencing: the combination of "being physically trapped" in front of the screen and "the cognitive load from producing and interpreting nonverbal cues" (Fauville et al., 2021b) makes referencing a common context and creating shared attention and connection difficult. Academic classrooms and workplaces, marked by a high frequency and intensity of video conferencing, were shown to exacerbate Zoom fatigue, as did factors such as lower economic status, poor academic performance and unstable internet connections (Oducado et al., 2021).

Yet video conferencing users need not wait for technological upgrades to reduce their fatigue. Bennett et. al. (2021) offer a number of recommendations for reducing Zoom fatigue, as illustrated in Table 2. Solid recommendations include better meeting times, improved group belongingness and muting microphones when not speaking; less certain recommendations include turning off webcams, using "hide self" view, taking breaks, and establishing group norms.

**Table 2.** *A table explaining recommendations for reducing video conference fatigue, adapted from figure 6 of Bennett et. al.'s study on video conference fatigue (2021).*

| Recommendations supported by our quantitative study our quantitative study | Potential explanation for fatigue reduction |
| --- | --- |
| 1. Hold meetings at a time that is least fatiguing for as many participants as possible based on work schedule, which may be earlier in the work period | Meetings are affect-generating events that may influence fatigue trajectory over the course of a day. |
| 2. Enhance perceptions of group belongingness | Enhanced perception of belongingness is expected to encourage interest in participation, reducing effortful attention and fatigue. |
| 3. Unless you are speaking, mute your microphone | Muting reduces both the potential for distracting background noise and the amount of active attention to stay quiet on the user's part. |
| **Recommendations with inconclusive evidence from our quantitative study** | **Potential explanation for fatigue reduction** |
| 4. Decrease/increase webcam usage | Increased webcam usage may increase group belongingness (and reduce fatigue), while decreased usage decreases stimuli and allows detaching, also possibly reducing fatigue. |
| 5. Consider using "hide self" view | Hiding the self camera potentially reduces stimuli and how much users worry about their appearance/background, improving belongingness. |

| Recommendations based on qualitative comments | Potential explanation for fatigue reduction |
|---|---|
| 6. Take breaks during videoconferences and between videoconferences | Breaks between and/or during meetings allow users to detach, a key method of reducing fatigue. |
| 7. Establish group norms (e.g., usage of mute and webcam, acceptability of multitasking, when/how to speak up) | Strong norms reduce ambiguity about acceptable behavior, and reduces active worry that contributes to fatigue. Additionally, they increase group belongingness. |

Zoom fatigue is a constellation of many different communicative problems with current videoconferencing. While Table 2's recommendations are a solid step in the right direction, they do not address the conferencing technology itself, a necessary step to begin reducing the need for users to compensate for the technology's shortcomings. Video conferencing research long predates Zoom and has also identified shortcomings, devised and applied methods for measuring communicative success, and proposed potential solutions. Below, we review these shortcomings, measures and solutions.

## 4 SHORTCOMINGS OF VIDEO CONFERENCING

Video conferencing has been with us for nearly a century (Peters, 1938), and research on its limitations predates the pandemic by decades. While research on video conferencing's long-term effects is sparse, researchers did investigate many of the same shortcomings studied by Zoom fatigue investigators such as Riedl et. al. (2021). We group the most relevant work into projects addressing problems with delay, gaze, objects of discussion, and a variety of non-verbal conversational cues.

**4.1 Delay**

Delay (lag) remains one of the most widely researched of video conferencing's technical shortcomings (Table 1, Shortcomings, Delay column), and represented in 12 out of 21 papers within Table 1's Measures, technical column. Delay for video conferencing is defined as the time elapsed between the moment of input (e.g. a joke or a smile) and the resulting response (e.g. a retort or a laugh) — "glass-to-glass." As a factor often outside of the control of users, the majority of research compares delay's quantitative severity against its qualitative effects. VideoLat is one system for measuring glass-to-glass video conferencing delays (Jansen and Bulterman, 2013). Users display a QR code to the camera, and compare the times that it is detected by the camera and displayed on an output monitor. We discuss more tools for measuring delay in Section 5.1.

A number of studies defined "significant" delay as 500-650ms (Schmitt et al., 2014; Whittaker, 2002; Tam et al., 2012). At such levels, delay causes "prolonged overlap, gap and sequential disarray and missed attempts at turn-taking" (see Table 3) (Olbertz-Siitonen, 2015), in addition to increased interruptions in video settings (O'Malley et al., 1996), all of which contribute to lower conversational quality. Schoenenberg (2016) defined a Quality of Mediated Conversation measure, composed of "the Conversational Quality, the Mediated Interaction, the Experiencer, the Interaction Partners and the Circumstances". Even one active user with significant delay can negatively impact the entire group's Quality of Experience (QoE), with the QoE decreasing as the delay becomes more symmetrical (equally distributed across participants) (Schmitt et al., 2014). Additionally, Becher et. al. (2020) found that while communicating collaboratively in an immersive virtual reality environment, increasing added delay from 300 ms to 450 ms introduced a noticeable decrease in mutual understanding, alongside a consistent decrease in task performance. Su et al. (2014) attempted to mask delay with prerecorded video or predicted motion, and found that masking was effective up until 800ms, but frequent masking was still required at 200ms.

**Table 3.** *A table illustrating turn-taking in a collaborative puzzle task, adapted from Gergle et. al.'s study (2012).*

| Turn | User A/ User B | Dialogue/[Action] |
|---|---|---|
| 1 | A | alright, um take the main black one |
| 2 | A | and stick it in the middle |
| 3 | B | [moves and places correct piece] |
| 4 | A | take the one-stripe yellow |
| 5 | A | and put it on the left side |
| 6 | B | [moves and places correct piece] |
| 7 | A | uh yeah, that's good |
| 8 | A | take the um two stripe one |
| 9 | A | and put it on top of the black one |
| 10 | B | [moves and places correct piece] |
| 11 | A | and take the half shaded one |
| 12 | A | and put it diagonally above the one that you just moved to the right |
| 13 | B | [moves and places correct piece] |
| 14 | A | yup, done. |

In addition to disrupting conversation and its general quality, video conferencing delay has emotional effects. Delays over 100ms have impacted user feelings of "fairness" in competitive events hosted on video conferencing, like a quiz game scenario (Ishibashi et al., 2006), with perceived fairness degrading as delay increases. Symmetrical and asymmetrical delay of 1200ms has caused users to mistakenly attribute technical issues to personal shortcomings (Schoenenberg et al., 2014; Schoenenberg, 2016), where users were likely to feel that delayed conversational partners were inattentive, undisciplined, or less friendly.

Because it is largely a technical problem, delay's effects are difficult for users to solve themselves. For example, reducing video quality may reduce delay, but it also reduces visual communicative cues. These studies confirm that even a fraction of a second (100-500ms) of delay can create conversational challenges. However, most of these studies examined only single video conferences. We suspect that, over several conferences (i.e. a fairly typical day of remote work), even less delay (<100ms) may suffice to increase the cognitive effort Riedl et. al. (2021) include in their model, and create Zoom fatigue.

**4.2 Gaze**

Gaze awareness is the ability to identify what — or importantly, who — a person is looking at. The majority of papers offering video conferencing solutions we reviewed (see Table 1's Focus/solution column) addressed gaze. These investigations are especially important, given few of today's common video conferencing systems can effectively depict gaze. In the real world, our view of a conversational partner and their view of us correspond. However in video conferencing systems, because camera and display are rarely colocated, this correspondence is broken. Even minor offsets in camera and display can notably affect our ability to recognize whether we are being looked at (Grayson and Monk, 2003). Gaze depiction becomes even more problematic as the number of conference participants grows: one camera cannot colocate with many participants. Yet when gaze can be effectively communicated by video conferencing, it has notable effects, especially in light of the importance of eye contact in mediated communication (Bohannon et al., 2013).

Among the papers addressing the challenges video conferencing systems have in communicating gaze (Table 1, Shortcomings, Gaze column), GA Display (see Figure 2) was among the earliest. An experimental video conferencing system supporting gaze awareness (Monk and Gale, 2002) for two users, GA Display used translucent screens and half-silvered mirrors. Its users were much more efficient in a conversational game (55% less turns and 949 less words) than users of an audio-only system. Multiview (Nguyen and Canny, 2005; Nguyen and Canny, 2007) supports two groups communicating with correct gaze through a shared virtual window. Multiview users formed trust relationships more quickly than users of traditional video conferencing systems. Even when collaborative tasks are not primarily communicative and video is not used, shared gaze awareness can be helpful. For example, gaze visualization has been shown to positively affect performance (D'Angelo and Gergle, 2018). When pair programmers are refactoring code, shared gaze awareness achieved with eye trackers and highlights showing gaze on shared in code helped increase task speed (D'Angelo and Begel, 2017).

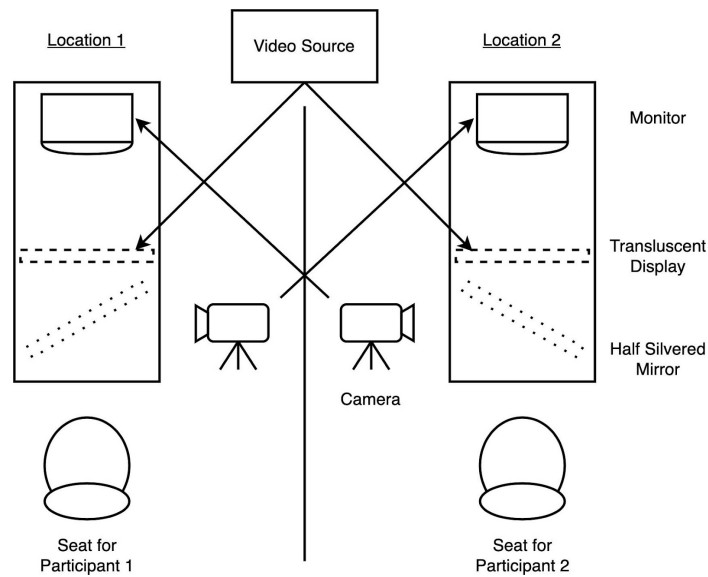

**Figure 2.** *An illustration of how GA Display creates gaze awareness in its video conferencing solution, adapted from Monk et. al.'s study (2002)*

Gaze's importance has resulted in a number of solutions in addition to GA Display and Multiview, which focus more on engineering the solutions themselves than on understanding the importance of gaze (Edelmann et al., 2013; He et al., 2021; Lawrence et al., 2021). We discuss these in more detail in Section 6. Gan et al. (2020) use ethnographic methods to study an underserved use of video conferencing, in which maintaining gaze is central: three-party video calls wherein one participant (e.g. a child) is less able to manage the technology, so that another (e.g. a grandparent) must help them speak with the third participant (e.g. a parent). They identify a range of needs for future video conferencing solutions to address.

Although few existing video conferencing solutions rely on it (e.g. (D'Angelo and Begel, 2017)), gaze tracking may play an important role in maintaining gaze awareness in the future. Fortunately, gaze tracking technology is already quite effective and quickly becoming more so: recent systems have achieved a refresh rate of 10000 Hz using less than 12 Mbits/s of bandwidth (Angelopoulos et al., 2021), or even power draws as low as 16 mW that are still accurate to within 2.67° while maintaining 400 Hz refresh rates (Li et al., 2020). Power and refresh rate concerns are especially important for XR headsets, in which power and latency can hinder not only eye-tracking effectiveness, but general comfort. Headsetless solutions to video conferencing will likely mandate sophisticated gaze tracking on top of other tracking technologies.

Like delay, lack of video conferencing support for gaze likely contributes to Zoom fatigue. As seen in Figure 1, lack of eye contact likely engenders a lack of shared attention, which in turn increases the cognitive load of video conferencing. Given the growing availability of inexpensive cameras and gaze tracking solutions, support for gaze may offer a widely applicable salve to long-term fatigue.

### 4.3 Objects of Discussion

Conversation often centers around a shared object, such as a whiteboard diagram, a presentation, or a working document. In such cases, discussion is filled with shorthand "deictic" references that make use of the object's context like "that one," "to the left of" and with pointing, "over there". Facilitating such conversational grounding has been particularly important for improving performance on shared tasks during video conferencing. Objects of discussion are an important part of shared attention and the papers we review (Table 1, Shortcomings, Object column), and a large portion of the papers that offer novel video conferencing solutions. With their ability to overlay virtual objects onto real world views, XR and AR are uniquely equipped to address this issue.

In two studies of communicative behavior, Gergle et al. (2012) and Kraut et. al. (2002) found that clear, synchronized and low delay shared visual information provides important feedback for successful communication. In more applied work, a novel system visualizing gaze onto code shared by pair programmers resulted in a notable improvement in performance of refactoring tasks across a range of metrics, including faster reference acknowledgement, more time with overlapping gaze and faster task completion (D'Angelo and Begel, 2017). ReMa (Feick et al., 2018), illustrated in Figure 3, tracks a user's manipulation of an object, and maps it to a remote robotic arm manipulating a proxy object. Initial evaluations showed that users preferred ReMa when combined with video conferencing, because it allowed a more intuitive understanding of shared artifacts. Initial studies in CollaboVR (He et al., 2020), a sketch based framework for collaboration in XR, found that among the projected, side-by-side and face-to-face (mirrored) layouts, users preferred face-to-face, citing that it allowed them to focus on both the shared artifact and their collaborator at the same time.

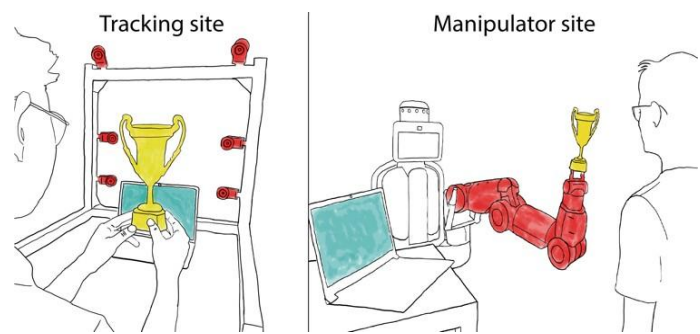

**Figure 3.** *An illustration of how Remote Manipulator operates, provided courtesy of Feick et. al. (2018)*

With their ability to overlay virtual objects onto real world views, XR and AR may be uniquely equipped to provide the grounding conversational context of objects of discussion. We could not find research examining support for objects of discussion in video conferencing over the long term, but expect that like reductions in delay and support for gaze, it may improve quality of communication and lower Zoom fatigue.

### 4.4 Additional Non-Verbal Cues

Beside gaze and objects of discussion, a variety of other non-verbal cues play an important role in conversation and are not commonly well-supported by existing conferencing systems, and include spatial location, gesture, facial expression and body language (Table 1, Shortcomings, Other Cues column). While none of these individual shortcomings is a dominant theme, collectively these shortcomings form a significant portion of the literature we review.

Gesture and expression are particularly important in speech formation and clarity, and have aided understanding in collaborative tasks (Driskell and Radtke, 2003; Driskell et al., 2003). Early studies confirmed the value of video in delivering such non-verbal cues, with video conferencing supporting more natural conversations as measured by improved turn taking, distinguishing among speakers, and better ability to interrupt/interject (Whittaker, 2002). When compared with audio-only solutions, video conferencing not only increased perceived naturalness, but also mitigated the impact of delays up to 500ms (Tam et al., 2012). Emphasizing the importance of visuals, Berndtsson et al. (2012) found that it was more important to synchronize audio and video than to reduce audio delays. This was true even at delays lower than 600ms (Berndtsson et al., 2012), with users preferring video conferencing over audio-only communication. In a virtual reality system without video, the introduction of more facially expressive avatars not only increased presence and social attraction, but also increased task performance (Wu et al., 2021).

Not all non-verbal conversational cues are visual. When improving the domestic video conferencing experience, Jansen et. al. (2011) found that spatial audio coupled with spatial audiovisual layout were necessary additions. With these features, communicating groups could more easily attend to central conversation, and ignore distractions in busy family environments.

Together with gaze and objects of discussion, these non-verbal cues form a suite of social markers that users cannot rely upon in common video conferencing solutions, increasing the cognitive effort and stress that form the backbone of long term fatigue. The work showing that such non-verbal cues can compensate for delays makes them a particularly promising avenue for future video conferencing solutions. XR technology should be particularly helpful in communicating location and gesture.

## 5 MEASURING VIDEO CONFERENCING EFFECTIVENESS

Any attempt at addressing video conferencing's shortcomings should be evaluated, to determine how well it supports human communication. While human-computer interface researchers know evaluative methods well, most are likely unfamiliar with evaluating communication itself. In this section, we review the methods previously used to evaluate video conferencing systems, with particular attention to those assessing communication. These fall into three categories: technical, behavioral and subjective (Table 1, Measures columns).

### 5.1 Technical Measures

Technical measures are those concerned with the performance of a system, and do not typically depend on human users. Such measures are most often used by researchers working with a systems focus, creating novel or improved solutions for video conferencing. The most commonly used technical measure is delay (Edelmann et al., 2013; Su et al., 2014; Gunkel et al., 2015; Schmitt et al., 2014), and related measures such as loss rates, audio and video quality (Edelmann et al., 2013; Jansen et al., 2011). Researchers often use delay in concert with other behavioral and subjective measures (O'Malley et al., 1996; Gunkel et al., 2015; Homaeian et al., 2021; Berndtsson et al., 2012).

Video conferencing researchers use many tools for measuring delay, including videoLat (Jansen and Bulterman, 2013), which measures the time elapsed between appearance on camera and on remote display ("glass-to-glass"); and vDelay (Boyaci et al., 2009), which measures delay similarly. Virtual reality researchers are also very concerned with delay, and often create communicative applications. Friston and Steed (2014) review methods of measuring latency, and describe a simple method for measuring delay, Automated Frame Counting, that makes use of a high-frame rate video camera. These tools are particularly useful for creating independent, consistent measurements of delay across varied, complex and sometimes closed video conferencing systems. A common element of these delay measuring methods is using camera footage or visual information (Jansen and Bulterman, 2013; Friston and Steed, 2014). For example, Roberts et al. (2009) compare communicative VR systems to video conferencing, noting that VR was much more effective at communicating attention, but had three times the delay of video conferencing (at 150ms).

## 5.2 Behavioral Measures

Like most computing systems, video conferencing is a tool that supports tasks, so conferencing effectiveness is often measured through its impact on tasks. These tasks can be simple conversation, or more applied work requiring informational exchange. Video conferencing is meant for multiple users, so tasks are often collaborative, and include finding a point on a shared object (Monk and Gale, 2002), word games (Driskell and Radtke, 2003), market trading (Nguyen and Canny, 2007), puzzle solving (Gergle et al., 2012), code refactoring (D'Angelo and Begel, 2017) and charades games (Wu et al., 2021). The goal of the evaluation is to measure the impact of a system improvement or experimental manipulation on task performance, either quantitatively or qualitatively.

### 5.2.1 Quantitative Measures

Traditional measures of task performance focus on efficiency. Video conferencing researchers make widespread use of both time (Wu et al., 2021; Monk and Gale, 2002; Gergle et al., 2012; O'Malley et al., 1996; Bennett et al., 2021; Homaeian et al., 2021) and accuracy (O'Malley et al., 1996; Monk and Gale, 2002; Driskell and Radtke, 2003; Wu et al., 2021). Other quantitative measures more specific to video conferencing include gaze estimation (Grayson and Monk, 2003) and counting gaze overlap with eye tracking (D'Angelo and Begel, 2017).

### 5.2.2 Communicative Measures

Communication researchers have devised a number of methods for assessing conversational efficiency and fluency, and these have naturally found application in studies of communicative systems like video conferencing. Typically, these communicative assessments involve recording the conversations on the system, then compiling various characterizing statistics. These include:

- *Word count:* a simple count of the number of words used in the conversation (O'Malley et al., 1996; Monk and Gale, 2002). Fewer words imply more efficient conversation (and better video conferencing).

- *Turn count:* as illustrated by Table 3, conversations can be parsed into a series of "turns," with each participant successively responding to what the other has said. Researchers perform this parsing, then count the number of turns (Monk and Gale, 2002; Gergle et al., 2012; O'Malley et al., 1996; Olbertz-Siitonen, 2015). Fewer turns are a more direct measure of conversational efficiency than word count.
- *Interruption count:* a count of when two or more participants are speaking at once (Monk and Gale, 2002; O'Malley et al., 1996; Schoenenberg, 2016; Tam et al., 2012; Olbertz-Siitonen, 2015). Fewer interruptions indicate more conversational efficiency.
- *Pause count:* a count of when no participants are speaking for a significant length of time (Monk and Gale, 2002; Olbertz-Siitonen, 2015).
- *Deictic word count:* a count of the number of words that rely on context, typically provided by a grounding object of discussion (D'Angelo and Begel, 2017).

Homaeian et al. (2021) have also recently proposed a methodology for detailed analysis of conversational grounding, utilizing a diagramming system they developed called "Joint Action Storyboards". With this scheme, they can measure the relationship of user interfaces, user interaction and cognition during communicative grounding.

## 5.3 Subjective Measures

In contrast to technical and behavioral measures, subjective measures obtain direct feedback from users, sometimes in the form of interviews (Nardi and Whittaker, 2001), but more often as surveys. For example, the International Telecommunications Union (ITU) advocates measuring Quality of Experience (QoE), which is "The overall acceptability of an application or service, as perceived subjectively by the end-user" (ITU, 2007). When possible, it is usually best to use standard surveys, since they are tried and tested, easily compared, and do not require the effort of generating a bespoke survey. Standard surveys used in video conferencing research include:

- *ITU-R BT.500:* The ITU (ITU, 2000) describes very detailed procedures for measuring subjective quality of audio and visuals, culminating in a survey. Typically these are relatively short lists of closed questions addressing fidelity and overall experience. Examples can be found in several papers (Ishibashi et al., 2006; Berndtsson et al., 2012; Schmitt et al., 2014; Gunkel et al., 2015; Schoenenberg et al., 2014).
- *Trust:* if users trust other conferencing participants (Butler Jr, 1991), used in (Bos et al., 2002; Nguyen and Canny, 2007).
- *Group Belongingness:* whether one feels part of the conversational group (Kraut et al., 1998), used in (Bennett et al., 2021).
- *Interpersonal Attraction:* indicates liking and attraction of conversational partners (Oh et al., 2016), used in (Wu et al., 2021).
- *Social Presence:* to capture the sense that users are connected with others through the system (Nowak and Biocca, 2003), used in (Wu et al., 2021).
- *Copresence:* for assessing the feeling that the user is with other entities (Nowak and Biocca, 2003), used in (Wu et al., 2021).
- *Temple Presence Inventory:* for user presence when engaging with media (Lombard et al., 2009), used in (He et al., 2021).

Despite the advantages of standardized assessments, many studies create bespoke surveys for their own purposes, often driven by the specific needs of their research. For example, Tam et al. (2012) created a survey capturing "naturalness" of conversation.

# 6 VIDEO CONFERENCING SOLUTIONS

Over the years of research in video conferencing, many improvements have been suggested and prototyped. We split these into two categories, window-based solutions aimed at gaze, the most dominant solution target in Table 1's Solutions column; and solutions addressing other nonverbal cues that are poorly supported in video conferencing environments. By reviewing such solutions, we can build on them for future work, and identify gaps that are opportunities for further work.

## 6.1 Window-Based Solutions for Gaze

These video conferencing solutions seek to restore gaze and other spatial cues by defining a virtual window shared by conference participants, with two screens at different locations representing opposite sides of the window. As we have already discussed in Section 4.2, increased gaze awareness can increase productivity/efficiency (Monk and Gale, 2002; Koboyashi et. al., 2021), trust formation (Nguyen and Canny, 2005), and presence (Lawrence et. al., 2021). Modern solutions are more likely to utilize XR (Edelmann et al., 2013; Koboyashi et. al., 2021).

GA Display (Monk and Gale, 2002; see also Section 4.2) utilizes cameras pointed at half silvered mirrors to capture the gaze of two participants, and a translucent display in front of a monitor in order to display captured gaze over a shared object of discussion. With GA Display, users required 55% fewer turns and 949 fewer words than in an audio-only system. While an early solution, GA Display's benefits show the — still today largely untapped — potential of restoring shared gaze awareness, particularly in collaborative tasks.

Gaze-2 (Vertegaal et al., 2003) conveys gaze in small group conferences. It differs somewhat from the other solutions in this section in that it can serve three or more remote locations (not just two remote screens defining the two sides of a virtual window). Each user sees the videos of other users in a row of tiles. Eye trackers for each user determine who they are looking at, and this information is used to rotate the user's tile toward that other user in each user's view — just as people might rotate their heads. Gaze-2 also employs the eye tracker to choose the camera with least parallax from among several cameras pointed at each user, so that users appear to be looking "straight out" of their tile. Under testing, automated camera shifts didn't affect perception of eye contact, and weren't considered highly distracting. Gaze-2 is unique in supporting gaze outside of one-on-one conference settings, and is a worthy vein for further research, given the nearly two decades of subsequent advancement in video conferencing technology.

Multiview (Nguyen and Canny, 2005; Nguyen and Canny, 2007) allows not just two participants, but two small groups to converse through a shared virtual window. Each group has one screen, displaying the virtual window. The screen is retroreflective like traffic signs, reflecting light primarily back in the direction from which it arrived. For each participant, there is a matching camera and projector. A participant's projector is located close to their head, and with the retroreflective screen, ensures that each participant sees a unique display. A participant's camera is also located "close" to their head on the shared virtual window (e.g. for the rightmost participant, at the rightmost location on the screen) — but at the remote location, giving them a view of the remote site that closely approximates their view through the window. Under testing, Multiview users formed trusting relationships just as quickly as face-to-face conference participants.

Kuster et. al. (2012) restore mutual gaze in two-way conferencing by using image warping to colocate the camera and the display. In this way, when the user looks at their conference partner, the partner sees the user looking at them. Tracking and rendering are performed with a consumer GPU and Kinect sensor. This system is unique in working within traditional, single-camera conferencing systems, but the views synthesized with its image warping were not completely convincing.

GazeChat (He et al., 2021) is a novel audio-only conferencing system that represents users as gaze-aware 3D profile pictures — the eyes in the pictures moved, providing meaningful cues about where the corresponding conversational partner was looking, facilitated by webcam-based eye tracking and neural network rendering. In a 16-person study, GazeChat outperformed simple audio and video in feelings of eye contact, while significantly outperforming audio in user engagement.

### 6.1.1 XR Window-Based Solutions

More modern window-based solutions supporting gaze have included tracking, 3D display, and spatial audio to create a shared 3D space — a hallmark of XR technology.

Face2Face (Edelmann et al., 2013) might be viewed as an improved version of GA Display, adding a holographic projection screen supporting 3D viewing, a more compact and flexible form factor supporting a wider range of views, and touch interaction.

Kobayashi et. al.'s (2021) system improves gaze in two-way conferencing using a unique embedding of multiple cameras into a screen, rather than with Kuster et al.'s (2012) image warping. Quantitative testing shows that users can more accurately estimate gaze than they would in a single camera system. In the long run, we expect such systems to be compelling solutions, but significant engineering challenges remain to realizing time, particularly in inexpensive consumer systems.

Google's Project Starline (Lawrence et al., 2021) creates a rich virtual window supporting continuous view change, coupled with spatial audio and stereoscopic display. Rather than relying only on several discrete cameras, Starline uses a combination of high resolution 3D capture and rendering subsystems. Based on participant surveys, Starline is notably superior to standard video conferencing in creating presence, attentiveness, reaction-gauging and engagement.

## 6.2 Solutions Addressing Other Non-Verbal Cues

Other solutions prioritize support for non-verbal communicative cues other than gaze. including objects of discussion, location and gesture.

For example, Jansen et al.'s (2011) system is specifically for noisy, group-to-group calls from home. The system consists of a hidden microphone array for spatial audio, and a number of HD cameras to facilitate dynamic composition based on group movement. When compared to traditional conferencing solutions, Jansen et al.'s system offers more flexibility in the kind of tasks remote groups can engage in, such as playing networked digital games.

### 6.2.1 XR Solutions

However, most of these solutions make extensive use of XR technology. Consider MultiStage (Su et al., 2014), which is designed largely for stage performers, allowing remote performers to interact through a CAVE-like system on a virtual stage. Stages are equipped with sensors to detect actors and large displays visualizing the connected performance. MultiStage's primary innovation is a method of masking delays that can replace high latency actors with either pre-recorded video of said actor or a computable model.

ReMa or Remote Manipulator (Feick et al., 2018; see also Section 4.3) offers rich, six-degree-of-freedom conversational grounding (in location and orientation). When a user demonstrates a physical object to a remote partner by adjusting the object's orientation, ReMa will replicate this manipulation for the remote partner with a robotic arm and a duplicate of the object. This is accomplished with a combination of tracking equipment following the user with the original object, and their manipulations being mapped onto a robotic arm with a proxy object. Evaluation finds that ReMa served as an effective collaboration tool, especially when combined with video conferencing.

Wu et. al. built a camera-based tracking system for VR that captures additional non-verbal communication cues like body language and facial expressions (Wu et al., 2021), and maps them onto virtual avatars. The system bolsters collaborative VR environments by allowing a broader range of socialization. Compared against a VR system with body-worn trackers in a game of charades, the expressive system facilitated more social presence and attractiveness, and improved task performance.

CollaboVR (He et al., 2020; see also Section 4.3) is a framework for VR collaboration in a shared 3D environment, allowing for sharing of freehand 2D sketches, which can be converted into 3D models with procedural, realtime animations. Based on cloud architecture to reduce client-side computational load, it also offers side-by-side (integrated), face-to-face (mirrored), and projected layouts to reduce clutter. Studies showed that the face-to-face layout was preferred, as it minimized obstructions from others, while also allowing users to focus on their collaborator's response.

Yu et. al. (2021) created a 3D telepresence system that allows AR, MR and VR interactions in a shared 3D environment. Remote users, joining the scene in VR, are presented as 3D avatars, while local users were presented as either avatars, or a point cloud representation, that captured their entire bodies, although their upper face was obscured by their headset. Through a study of a teleconsultation task, the point cloud representation proved more effective, as users found it more expressive than the avatar, despite the obfuscation of upper facial details caused by point display.

While this list of conferencing solutions may seem extensive, we find it surprisingly short, given the decades of research on this topic, and especially the new importance of video conferencing. We believe many research opportunities remain, which we detail in Section 7 below.

## 7 OVERVIEW AND FUTURE RESEARCH

Table 1 summarizes the body of research we have reviewed, describing efforts to understand Zoom fatigue, human communication during video conferencing, deficits of current video conferencing technology, and proposed solutions. Over the course of our investigation, the importance of reducing the cognitive effort of conferencing by more effectively capturing and displaying aspects of in-person interaction has become evident. Not only do improvements upon communication improve productivity, they reduce the long-term strain of video conferencing When updated with modern XR technology, many of the solutions we surveyed may prove much more effective.

With this overview of video conferencing research, we can identify a wide range of scientific and engineering opportunities that remain underexplored:

**7.1 Opportunities for Scientific Research**

- *Zoom Fatigue:* Nearly all the research we surveyed studied single video conferences, rather than long-term conferencing across several remote meetings. Today this is a very common scenario that deserves attention. How important are reducing delay, communicating gaze, and offering objects of discussion in this long-term context? Answers could reprioritize work on engineered video conferencing solutions.

- *Gaze:* As video conferencing has become more commonplace, so have larger conferences with many participants. We expect that gaze will gain importance with conference size, but research should confirm this.

- *Delay:* Further research on delay in the modern context is needed. Much of the existing research is quite old, predating such technologies such as GPUs and machine learning, and applications such as widespread non-business use, gaming and large-scale teaching. Delays well under 100ms are important and possible in many other settings; it would not be unreasonable to find similar needs for conferencing, especially over the long term. Finally, little is known about the effects of variation in delay on communication.

- *Non-verbal cues:* This category of video conferencing shortcomings hides many that have seen little or no examination. In particular, we are not aware of any research on "big face" or backchanneling, and very little on gestures.

- *Neurological measures:* While we intended to include a column for neurological measures in Table 1, we found no research using these measures. This is troubling, since recent research shows that the human brain responds strongly to faces and social interaction (Hoehl et al., 2008). Further work should address this deficit as soon as possible, and may reveal phenomena otherwise missed in prior work.

- *Communicative measures:* Space did not allow us to break out these measures from their parent behavioral category, though we are confident that these measures are not being used enough to evaluate new video conferencing solutions. This should change, so that future engineering efforts can be more effectively evaluated.

- *Complex models of video conferencing:* While a great deal of research has investigated how one shortcoming affects video conferencing, we are aware of no research that studies how they interact. Consider Riedl et al.'s (2021) posited model of Zoom fatigue in Section 3: how strong are the relationships it depicts? Which are strong, which are weak? Answers to such questions will help prioritize future research.

**7.2 Opportunities for Engineering Research**

- *Overcoming shortcomings:* While much work has been done to overcome video conferencing's shortcomings, much remains. For example, how can some of the solutions for delivering gaze be delivered with typical, modest conferencing hardware? Are there ways of compensating for delay that do not introduce serious communicative tradeoffs? How might modern XR technologies be leveraged to improve previous solutions?

- *Video conferencing at scale:* As we have just mentioned when discussing gaze, modern video conferences commonly have dozens of participants. Even outside of gaze, little of the research we found addressed conferencing at these scales. This is unfortunate, because when conferencing takes place at this scale, it is at its worst, and the need for improvement is greatest. How can conferencing support grounding, recognition, interruption, and discussion at such scales? Researchers may find inspiration in the different types of meetings and purposes that real-life conferences support.

- *Heterogeneous video conferencing systems:* XR technologies are still emerging, and will not be ubiquitous for many years at least. Heterogeneous systems, with different technologies used by participants in the same conference, will be commonplace (e.g., Telelife (Orlosky et al., 2021)). How can the technical, health and social asymmetries of hybrid systems be accommodated, particularly in educational environments? A few studies of these issues exist (Yoshimura and Borst, 2021; Hopkins and Benford, 1998), but more are necessary to establish a complete picture of the complex effects on fatigue, fairness and diversity of such heterogeneity in conferencing technology.

- *Better-than-real conferencing:* Lastly, most research strives to make video conferencing as good as face-to-face. But where might it be better? For example, could video conferencing systems keep meetings effectively summarized and on schedule? Could they permit freedom of motion? Might they support design review more effectively than in-person meetings?

# 8 CONCLUSION

This paper has reviewed video conferencing's shortcomings, ways of measuring them, and attempts at addressing them — with an eye towards XR's potential impact on that direction. We paid particular attention to the ways that the legacy of conferencing research could apply to the recently prioritized issue of long-term Zoom fatigue. Despite the relative recency of the fatigue phenomenon prior studies and solutions should be applicable, if modernized and adapted to current standards.

Additionally, we revealed many remaining scientific and engineering research opportunities, including research employing neurological and communicative measures, which should guide future investigation; and video conferencing at scale. We hope that the next such research review will find that many of our remaining research questions will have at least initial answers.

## ACKNOWLEDGMENTS

Our sincerest thanks to Chung-Che Hsao and Professor David Berube for many fruitful conversations about video conferencing. This work was supported by North Carolina State University's Department of Computer Science.